# The Affects of Different Queuing Algorithms within the Router on QoS VoIP application Using OPNET

Dr. Hussein A. Mohammed*, Dr. Adnan Hussein Ali**,
Hawraa Jassim  Mohammed*

\* Iraqi Commission for Computers and Informatics/Informatics Institute
for Postgraduate Studies.
\*\* Electrical & Electronics Techniques College – Baghdad
(adnan_h_ali@yahoo.com)

**Abstract:**

   *Voice over Internet Protocol (VoIP) is a service from the internet services that allows users to communicate with each other. Quality of Service (QoS) is very sensitive to delay so that VOIP needs it. The objective of this research is study the effect of different queuing algorithms within the router on VoIP QoS.*

   *In this work, simulation tool "OPNET Modeler version 14.0" is used to implement the proposed network (VoIP Network). The proposed network is a private network for a company that has two locations located at two different countries around the world in order to simulate the communications within the same location as a local and the communications between two locations as a long distance and analyze VoIP QoS through measuring the major factors that affect the QoS for VoIP according to international telecommunication union (ITU) standards such as: delay, jitter and packet loss.*

   *In this research , a comparison was carried out between different queue algorithms like First in First out (FIFO), Priority queue (PQ) and Weight Fair Queuing (WFQ) and it was found that PQ and WFQ algorithms are the most appropriate to improve VoIP QoS .*

**Keywords:** *OPNET, FIFO Queue, Priority Queue, Weighted-Fair Queue, QoS, ToS.*

## 1.  INTRODUCTION

   Voice over IP (VoIP) uses the Internet Protocol (IP) to transmit voice as packets over an IP network, rather than the traditional telephone landline system which called Public Switched Telephone Network (PSTN).  VoIP can be achieved on any data network that uses IP, for example Internet and Local Area Networks (LAN). By VOIP the voice signal firstly digitized, compressed and converted to IP packets and after that it will be transmitted over the IP network. With this technology, the potential for very low-cost or free voice can be achieved. The increase in capacity of the Internet in addition to popularity gives an increasing need to provide real-time voice and video services to the network  [1].

## 2.  VoIP Quality of Service (QoS)

   QoS can be defined as the ability of the network to support good services in order to accept good customers. In other words, QoS measures to the degree of user satisfactions and network performance. Applications like FTP, HTTP, video conferencing  and e-mail are not sensitive to delay of transmitted information assess QoS in important problems, while other applications like voice and video are more sensitive to loss, delay and jitter of the information. Therefore, QoS of VoIP is an import interest to ensure that voice packets are not delayed or lost while be transmitted over the network [2].

VoIP QoS is measured according ITU recommendations based on different parameters like (delay, jitter, and packet loss), these parameters can be changed and controlled within the acceptable range to improved VoIP QoS. Factors affecting QoS are briefly described in the following sections [2]:

**1. Latency:** As a delay sensitive application, voice cannot tolerate too much delay. Latency is the average time it takes for a packet to travel from its source to its destination. A person whose speaking into the phone called the source and the destination is the listener at the other end. This is one-way latency [3]. Ideally, must keeping on the delay as low as possible but if there is too much traffic on the line (congestion), or if a voice packet gets stuck behind a bunch of data packets (such as an email attachment), the voice packet will be delayed to the point that the quality of the call is compromised [4]. The Maximum amount of latency that a voice call can tolerate one way is 150 Milliseconds (0.15 sec) but is preferred be 100 Milliseconds (0.10 sec) [5].

Equation (1) shows the calculation of delay where **Average delay (D)** is expressed as the sum of all delays (*di*), divided by the total number of all measurements (*N*) [6].

$$D \quad = \quad \sum_{i=1}^{N} \quad di \quad \Big/ \quad N$$

... *(1)*

**2. Jitter (Variation of Delay)**: In order for voice to be intelligible, voice packets must arrive at regular Intervals. Jitter describes the degree of fluctuation in packet access, which can be caused by too much traffic on the line [4]. Voice packets can tolerate only about 75 Milliseconds (0.075 sec) but is preferred be 40 Milliseconds (0.040 sec) of jitter delay [5].

Equation (2) shows the calculation of jitter (j). Both average delay and jitter are measured in seconds. Obviously, if all (*di*) delay values are equal, then *D = di* and *J = 0* (i.e., there is no jitter) [6].

$$J = \sqrt{\frac{1}{N-1} \sum_{i=1}^{N} (di - D)^2}$$

... *(2)*

**Packet loss**: is the term used to describe the packets that do not arrive at the intended destination that happened when a device (router, switch, and link) is overloaded and cannot accept any incoming data at a given moment [7]. Packets will be dropped during periods of network congestion. Voice traffic can tolerate less than a 3% loss of packets (1% is optimum) before callers feel at gaps in conversation [5].

Equation (3) shows the calculation of packet loss ratio defined as a ratio of the number of lost packets to the total number of transmitted packets Where N equals the total number of packets transmitted during a specific time period, and $N_L$ equals the number of packets lost during the same time period [6].

**Loss packets ratio = ($N_L$ / N) × 100%**                    ... *(3)*

## 3. Queuing Disciplines

When the network is designed to service widely varying types of traffic, there is a way to treat contention for network resources by queuing , and manages the available resources according to conditions  outlined by the network administrator. Each router, as part of the resource allocation mechanisms, must implement some queuing (algorithm) discipline that governs how packets are buffered while wait to be transmitted. Various queuing disciplines can be used to control the transmitted packets. The queuing disciplines also affects to the packet latency by decreasing the time that packets wait to be transmitted. There are three common queuing disciplines that can be analyses, they  are first-in-first-out (FIFO) queuing, priority queuing (PQ) and weighted-fair queuing (WFQ) [8].

The basic principle  of  FIFO queuing is that the first packet that arrives at a router is the first packet to be transmitted. An exception here happened if a packet arrives and the queue  is full, then the router ignores that packet at any conditions . [8]. As shown in the Figure 1.

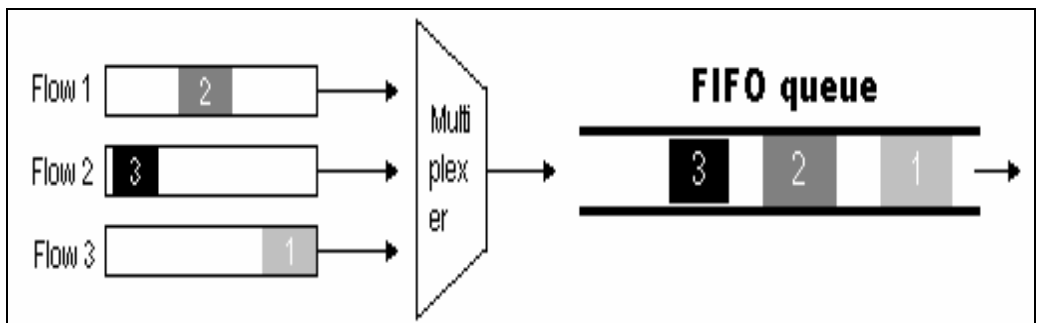

Figure 1: FIFO queue [11]

 The principle idea of PQ queuing depends on the priority of the packets, a highest priority are transmitted on the output port first and then  the packets with lower priority and so on. When congestion occurs, packets with lower-priority queues will be dropped . The only problem with these packets is that has lower-priority in queue [9]. As shown in the figure2.

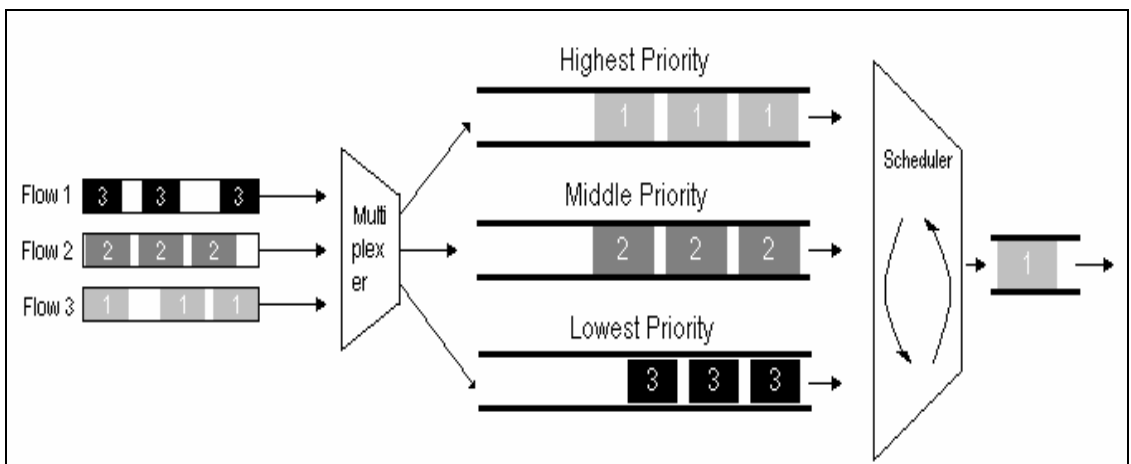

Figure 2: Priority queuing [11]

The Weighted-fair queuing discipline provides QoS by provides fair (dedicated) bandwidth to all network traffic for control on jitter, latency and packet loss. The packets are classified and placed into queues according to information ToS field in IP header is use to identify weight (bandwidth). The Weighted-fair queuing discipline weights traffic therefore a low-bandwidth traffic gets a high level of priority.  A unique feature of this queuing discipline is the  real-time interactive traffic will be moved  to the front of queues and fairly the other bandwidth shares among other flows [9]. As shown in the Figure3.

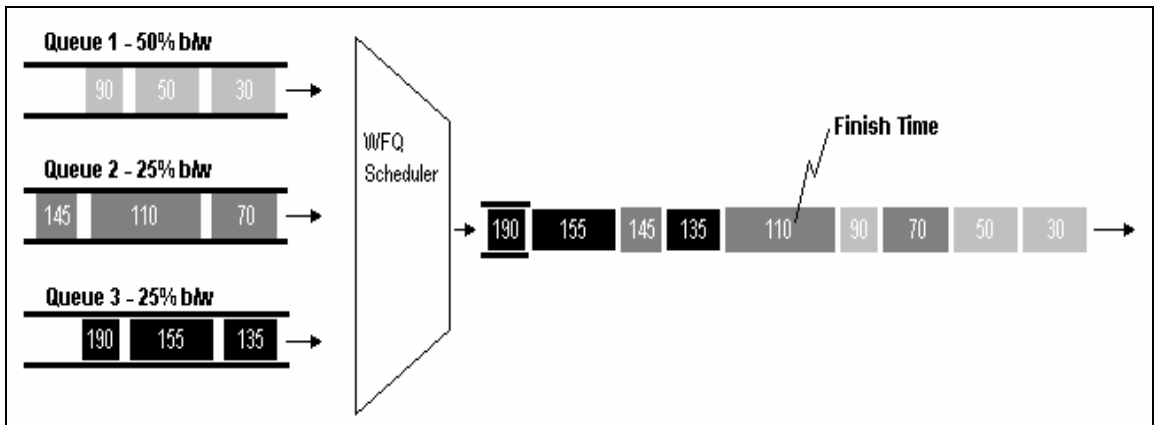

Figure 3: WFQ queuing [11]

# 4.  OPNET Simulator

The simulation tool adopted in this research is Optimized Network Engineering Tools (OPNET) version 14.0. OPNET is an object-orientated simulation tool for making network modeling and QoS analysis of simulation of network communication, network devices and protocols. OPNET Modeler has a vast number of models for network elements, and it has many different real-life network configuration capabilities. These make real-life network environment simulations in OPNET very close to reality and provide full phases of a study. OPNET also includes features such as comprehensive library of network protocols and models, user friendly GUI (Graphical User Interface), Web report is feature allows you to organize and distribute the results of your simulations in form graphical results and statistics [10]. OPNET doesn't  have any programming knowledge so that its easy to use and to deal with for any person [11]. Automatic simulation generation OPNET models can be compiled into executable code [7].

# 5.  Simulation work:

## Implementing the Proposed Network with OPNET

The network model for simulation consists from two companies that are located in two different countries around the world. The local area network (LAN) structure for both companies is the same (as shown in Figure 4).

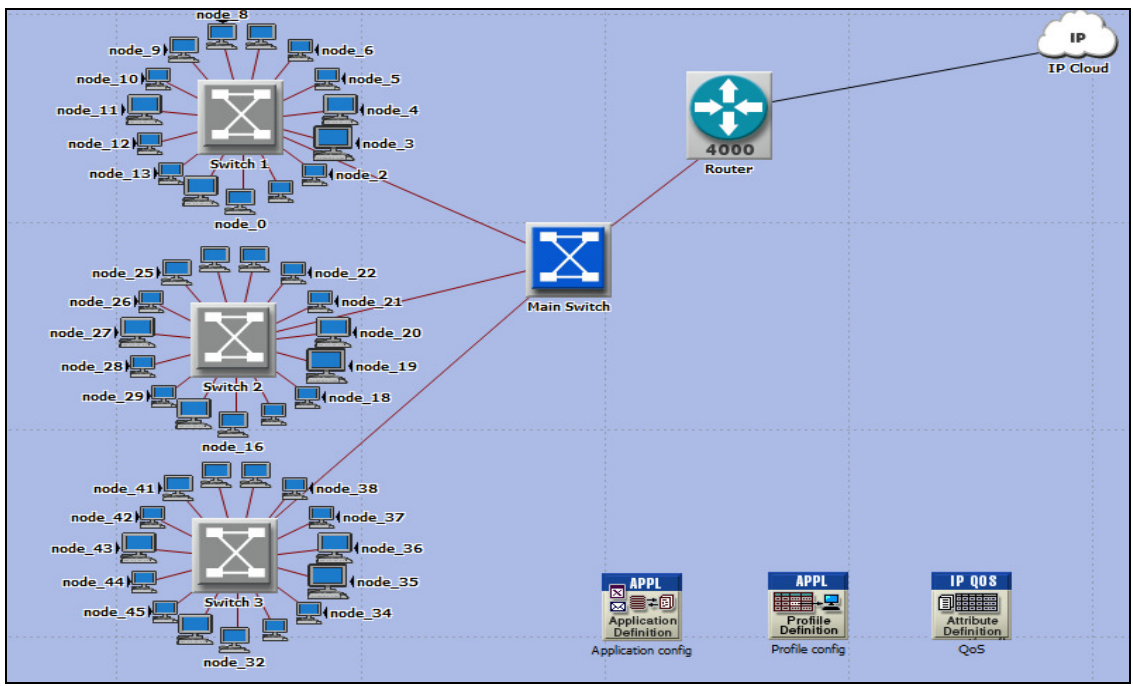

Figure 4: The Simulation Network Model.

In each LAN (three floors and in each floor are 15 Ethernet workstations, three Ethernet switch , one main switch called (Bay Networks Centilion100 Switch)  and  one  router called (Cisco C4000 Router) connecting to IP cloud by bi-directional PPP_DS1 link. The workstations are connected to the switches by bi-directional 10Base_T links. The proposed network carries applications FTP, Video simultaneously along with VoIP.

There are some important objects in OPNET that are used to simplify and facilitate the work with it. In this research, some of these objects that are used are Application Configuration, Profile Configuration and QoS Attribute Configuration as show in Fig.4:

**1. Application Configuration Object:** is an object used to define and configure all Applications in the network according to the user requirements. OPNET has most common applications like: HTTP, E-mail, video, File transfer, Voice, database. In this research, there is one application in the network model is a Voice.

**2. Profile Configuration Object:**  is an object that can be used to create users profiles, profile can contains one or more applications and each application can be configured by the starting time and ending time. In this research, there is only one profile to all users called VoIP_ profile.

**3. The QoS Attribute node:** is a mean of attribute configuration details that assess protocols at the IP layer. It deals with the three queuing profiles: FIFO, Priority queuing and Weighted-fair queuing.

# 6. Network Simulation Results and Discussions

In this research, a Comparison between the effects of different queuing duplicates such as FIFO, PQ and WFQ on VoIP QoS. To measure the QoS of the VoIP application during collected statistics (parameters) such as: Voice delay (sec), Voice jitter (sec), Voice traffic sends (packet/sec) and Voice traffic received (packet/sec). The duration of simulation is 8 minutes and the results are obtained as shown in figure5.

The blue line represent FIFO algorithm; whereas the red line represent PQ algorithm; whereas the green line represent WFQ algorithm.

***Figure (a)*** shows the (Jitter = 0) using PQ and WFQ algorithms; whereas jitter using FIFO algorithm = 0.0038 sec but did not exceed the time constraint (0.075 sec).

***Figure (b)*** shows the end to end delay using PQ and WFQ algorithms acceptable value but using FIFO algorithm did exceed the time constraint (0.15 sec).

***Figure (c)*** shows the voice traffic received using FIFO algorithm = 380 (Packets/sec), the voice traffic received using PQ and WFQ = 400 (Packets/sec).

***Figure (d)*** shows the voice traffic sent in all three cases = 400 Packets/sec.

Figure (5) shows that in FIFO queue, the delay does exceed the time constraint 150 ms and more packet loss. According to OPENT, packet loss ratio is the ratio of packets dropped to the total packets transferred to this cloud multiplied by 100%.

Packet loss ratio = 20/400×100% =5%.

In PQ and WFQ algorithms have no packet loss and end to end delay with acceptable ratio. Jitter in FIFO, PQ, WFQ has acceptable ratio.

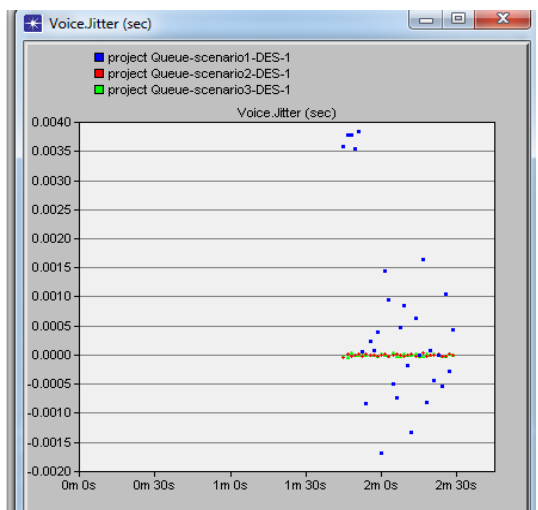

*(a) Voice Jitter*

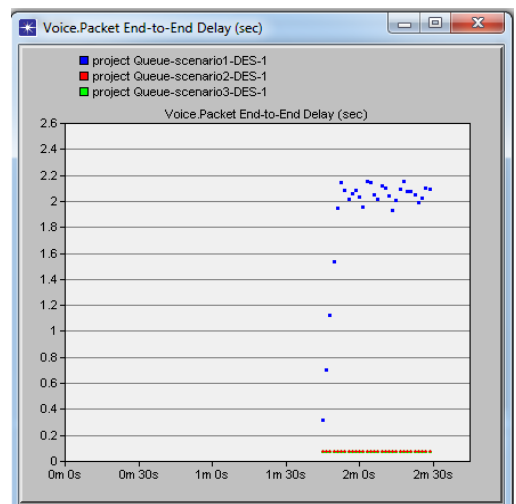

*b) Voice Packet end to end Delay*

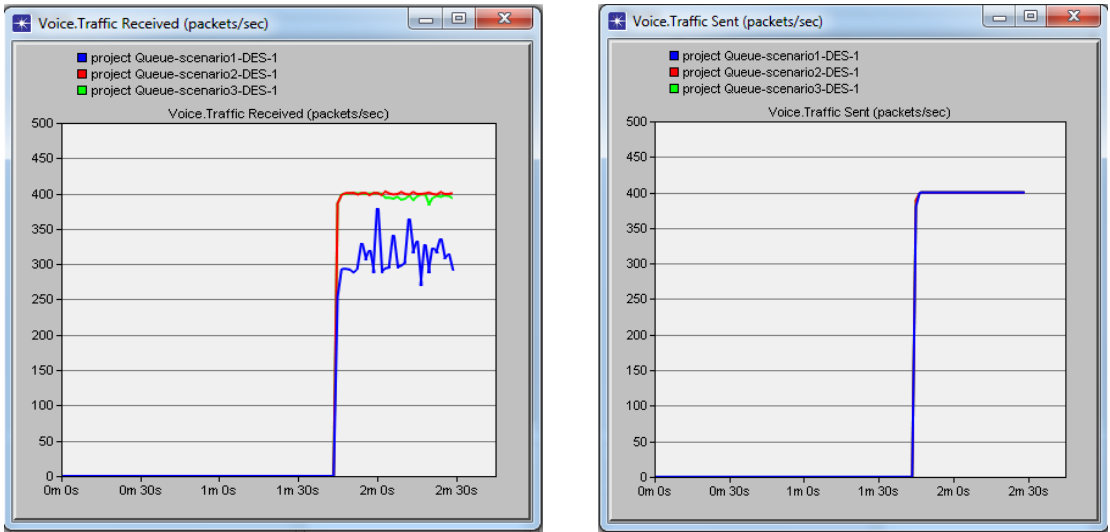

*(c) Voice Traffic Received (packet/sec)*        *(d) Voice Traffic Send (packet/sec)*

Figure 5: Comparison between Different Queuing Disciplines and Effect on VoIP QoS.

Different parameters captured to different queuing disciplines (FIFO, PQ and WFQ) are summarized in Table 1. From the result summarized in Table 1, The PQ and WFQ are the most appropriate scheduling schemes for improve QoS of voice traffic.

**Table 1: Results of Different Queuing Disciplines on VoIP Quality.**

| Parameters | FIFO | PQ | WFQ |
|---|---|---|---|
| Voice Jitter (sec) | 0.00383 | 0.00003 | 0.00003 |
| Voice Packet End-to-End Delay (sec) | 2.15 | 0.076 | 0.076 |
| Voice Traffic Received (packets/sec) | 380 | 400 | 400 |
| Voice Traffic Sent (packets/sec) | 400 | 400 | 400 |
| Packet loss | 20 | 0.0 | 0.0 |

# 7.  Conclusions

The presented research regards with the affects of different queuing disciplines on the performance of VoIP using OPNET. Simulations results allow us to conclude that; Improving the QoS of voice traffic based on  the Priority and Weighted-fair queues are the most appropriate scheduling schemes because the values of the parameters are within the acceptable range such as delay, jitter, packet loss.

## Authors

Dr Adnan H. Ali   is currently a Ph D. degree in Laser and Opto-electronic Engineering from university of Technology-Baghdad, He is now with Computer Techniques Engineering   Department at Electrical & Electronics Techniques College - Baghdad.   ( adnan_h_ali@yahoo.com).

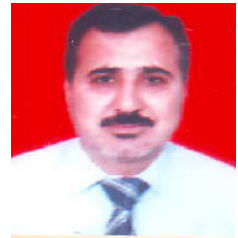